\documentclass{ifacconf}
\usepackage[usenames,dvipsnames]{xcolor}
\usepackage{graphicx}
\usepackage{natbib}
\usepackage{amssymb}
\usepackage{amsmath}
\usepackage{mathtools}
\usepackage{eurosym}
\usepackage{derivative}
\usepackage{threeparttable}
\usepackage{booktabs}
\usepackage{bm}
\usepackage[FIGTOPCAP]{subfigure}
\usepackage[short]{optidef}
\usepackage[version=3]{mhchem}
\usepackage{float}
\usepackage{dirtytalk}
\normalsize
\graphicspath{{figs/}}

\begin{document}
\begin{frontmatter}

\title{Switching-time bioprocess control with pulse-width-modulated optogenetics}

\author[First]{Sebastián Espinel-Ríos} 

\address[First]{School of Chemical and Bioprocess Engineering, University College Dublin, Ireland (e-mail: sebastian.espinelrios@ucd.ie)}

\begin{abstract} 
Biotechnology can benefit from dynamic control to improve production efficiency. In this context, optogenetics enables modulation of gene expression using light as an external input, allowing fine-tuning of protein levels to unlock dynamic metabolic control and regulation of cell growth. Optogenetic systems can be actuated by light intensity. However, relying solely on intensity-driven control (i.e., signal amplitude) may fail to properly tune optogenetic bioprocesses when the dose-response relationship (i.e., light intensity versus gene-expression strength) is steep. In these cases, tunability is effectively constrained to either fully active or fully repressed gene expression, with little intermediate regulation. Pulse-width modulation can alleviate this issue by alternating between fully ON and OFF light intensity within forcing periods, thereby smoothing the average response and enhancing process controllability. Optimizing pulse-width-modulated optogenetics entails a switching-time optimal control problem with a binary input over multiple forcing periods. While this can be formulated as a mixed-integer optimization problem on a refined control grid with monotonic input constraints, the number of decision variables can grow rapidly with increasing control-grid resolution within forcing periods and with the total number of forcing periods, complicating the task. Here, we propose an alternative solution based on reinforcement learning. We parametrize control actions via the duty cycle, a continuous proxy variable that encodes the ON-to-OFF switching time within each forcing period, thereby respecting the intrinsic binary nature of the light intensity while avoiding fine-grid binary decision variables.
\end{abstract} 

\end{frontmatter}
\section{Introduction}
Biotechnology leverages the enzymatic pathways of microorganisms to synthesize chemicals, materials, and fuels, among other valuable products. Microbial bioproduction therefore plays a crucial role in the development of a future circular economy \citep{ewing_fermentation_2022, konzock_trying_2024}. One emerging technology to optimize bioproduction is optogenetics, which can be used to fine-tune gene expression in real-time using light as an external control input \citep{milias-argeitis_automated_2016, hoffman_optogenetics_2022, benisch_unlocking_2024}. Either via gene expression activation or repression, this enables dynamic modulation of intracellular enzyme levels throughout bioprocess operation, for example, to achieve dynamic metabolic control. It also shows potential for regulating microbial growth through modulation of the expression of proteins involved in toxin-antitoxin systems, antibiotic-resistance modules, or auxotrophic amino acid synthesis. These capabilities promise enhanced control of population levels in synthetic microbial communities.

Optogenetic bioprocesses can be actuated via light intensity (i.e., a continuous signal amplitude) of specific wavelengths. Light-intensity-driven optogenetics has been explored through model-based predictive control and reinforcement learning (RL) (e.g., \cite{milias-argeitis_automated_2016, espinel-rios_reinforcement_2025, espinelrios_rl_metab_2025}). In an optimal control context, light intensity can be discretized over finite piecewise-constant control intervals that act as the degrees of freedom. However, intensity-driven optogenetic control can suffer from limited tunability and robustness \citep{davidson_programming_2013, benzinger_synthetic_2022}. In many optogenetic systems, dose-response curves (i.e., light intensity versus gene activation/repression) are very steep, driving the system close to fully active or fully repressed gene-expression states and thus hindering intermediate operating points.

Pulse-width modulation (PWM) is an alternative way to drive optogenetic bioprocesses \citep{davidson_programming_2013, benzinger_synthetic_2022} that can address the issue of steep dose-response. It exploits two light-intensity levels (ON/OFF) over predefined forcing periods (a full interval of ON and OFF subintervals). The ON state typically corresponds to the maximum achievable light intensity, and the OFF state to zero intensity. This strategy improves practical tunability and robustness by \emph{averaging} the response over the forcing period.

Optimizing PWM-driven bioprocesses is, however, nontrivial. The optimal solution of a PWM-driven optogenetic bioprocess entails optimizing the switching times for the many forcing periods throughout the process. This leads to a switching-time optimal control problem with a binary input over multiple forcing periods. One way to formulate this is as a \textit{mixed-integer optimization problem}, where time is finely discretized within each forcing period and constant binary inputs are optimized over the resulting subintervals while enforcing the constraint that once an input switches OFF, all remaining inputs in that period must remain OFF. However, this would require fine control-interval discretization to accurately approximate the continuous switching points. In addition, such formulations may increase the size of the optimization problem as the time grid becomes finer and the number of forcing periods grows. Coarser discretization may facilitate tractability but restricts the solution space. Here, we use RL with duty cycles as continuous proxy variables that encode binary light inputs, thereby avoiding fine-grid binary decision variables and making the formulation compatible with continuous-action learning frameworks. RL policies can be derived \textit{in silico}, for example, through interactions with a digital twin.

The remainder of this paper is structured as follows. Section \ref{sec:optimal_control_problem} formulates the switching-time optimal control problem underlying PWM. Section \ref{sec:solution_strategy} outlines our proposed strategy based on RL. Finally, Section \ref{sec:case_study} presents a case study involving optogenetic control of \textit{Escherichia coli} growth via PWM. Efficient control of cell growth is relevant for bioprocess applications involving microbial consortia, where different cell populations can carry out distinct metabolic reactions \citep{hoffman_optogenetics_2022, benisch_unlocking_2024}. Robustness of the RL training procedure is demonstrated via randomization of the dynamics and initial conditions.

\section{Switching-time optimal control}
\label{sec:optimal_control_problem}
In PWM-driven optogenetics, the optogenetic input $\bm{u}(t)$ is modeled as a binary-valued vector of dimension $n_u$, where $n_u$ is the number of inputs (i.e., independent light intensity channels). For each input channel $i$, we define:
\begin{equation}
    u_i(t) \in \{0, 1\},
\end{equation}
where $u_i(t) = 0$ (OFF) encodes $0\,\%$ and $u_i(t) = 1$ (ON) encodes $100\,\%$ of a given maximum light intensity $I_{i,\max} \in \mathbb{R}$. The corresponding \textit{physical} light intensity for channel $i$ is thus:
\begin{equation}
    I_i(u_i(t)) = u_i(t)\, I_{i,\max}.
    \label{eq:physical_i_I}
\end{equation}

Furthermore, the overall process involves $n_T$ forcing periods, so that $t_f = n_T T$ denotes the finite process time. We define the $k$-th forcing period $\mathcal{T}_k$ as the interval:
\begin{equation}
    \mathcal{T}_k := [kT,\,(k+1)T), 
    \quad \forall k \in \{0, 1, \dots, n_T-1\}.
\end{equation}

The trajectory of input $i$ over the $n_T$ forcing periods of the process follows a binary ON-OFF pattern within each forcing period:
\begin{equation}
\begin{aligned}
    u_{i,k}(t) &=
    \begin{cases}
        1, & t \in [kT,\, \tau_{i,k}),\\[3pt]
        0, & t \in [\tau_{i,k},\, (k+1)T),
    \end{cases}
    \\
    &\forall k \in \{0, 1, \dots, n_T-1\},
\end{aligned}
\label{eq:binary_input}
\end{equation}
where $kT \leq \tau_{i,k} \leq (k+1)T$ denotes the switching time at which the input transitions from ON to OFF within the period $\mathcal{T}_k$. In a PWM-driven optogenetic bioprocess, the decision variable within the period $\mathcal{T}_k$ is therefore the switching time $\tau_{i,k}$. Here we assume that each forcing period has fixed duration $T$, set \textit{a priori}. The boundary cases $\tau_{i,k}=kT$ and $\tau_{i,k}=(k+1)T$ are admissible and correspond to fully OFF and fully ON forcing periods, respectively.

Let us now consider an optogenetic bioprocess with state $\bm{x}(t) \in \mathbb{R}^{n_x}$ and measured output $\bm{y}(t) = \bm{h}(\bm{x}(t))$, where $\bm{h}: \mathbb{R}^{n_x} \mapsto \mathbb{R}^{n_y}$ captures the underlying measurement function. The continuous-time dynamics read:
\begin{equation}
    \odv{\bm{x}(t)}{t} = f\left(\bm{x}(t), \bm{I}(t)\right), 
    \quad \bm{x}(t_0) = \bm{x}_0,
    \label{eq:dynamics}
\end{equation}
where $f: \mathbb{R}^{n_x} \times \mathbb{R}^{n_u} \mapsto \mathbb{R}^{n_x}$ captures the underlying process dynamics, $\bm{I}(t) \in \mathbb{R}^{n_u}$ is the vector of physical light intensities, $t_0$ is the initial time, and $\bm{x}_0$ the initial state.

With these definitions in place, the generalized \emph{switching-time optimal control problem with a binary input} is:
\begin{mini!}
    {\{\tau_{i,k}\}_{i=1,\;k=0}^{n_u,\;n_T-1}}{\mathbb{E}[J(\cdot)],}%
    {\label{eq:sw-opt}}{}
    \addConstraint{\odv{\bm{x}(t)}{t} = f\left(\bm{x}(t), \bm{I}(t)\right)}
    \addConstraint{\bm{x}(t_0) = \bm{x}_0}
    \addConstraint{\bm{y}(t) = \bm{h}(\bm{x}(t))}
    \addConstraint{
        u_{i,k}(t) = 
        \begin{cases}
            1, \, t \in [kT,\, \tau_{i,k}),\\
            0, \, t \in [\tau_{i,k},\, (k+1)T),
        \end{cases}
    }
    \addConstraint{I_{i,k}(u_{i,k}(t)) = u_{i,k}(t)\, I_{i,\max}}
    \addConstraint{kT \leq \tau_{i,k} \leq (k+1)T}
    \addConstraint{\forall k \in \{0, 1, \dots, n_T-1\}} \notag
    \addConstraint{\forall i \in \{1, \dots, n_u\}.} \notag
\end{mini!}
Here, $\mathbb{E}[J(\cdot)]$ denotes the expectation of an optimization objective function $J(\cdot)$, and the decision variables are the switching times $\tau_{i,k}$ for all inputs across all forcing periods.

Without loss of generality, in this paper we consider reference 
tracking control:
\begin{equation}
    J = \int_{t_0}^{t_f} 
        \| \bm{y}(t) - \bm{r}(t) \|^2_{\mathbf{Q_s}} \, dt
    + \| \bm{y}(t_f) - \bm{r}(t_f) \|^2_{\mathbf{Q_t}},
    \label{eq:cost}
\end{equation}
where the reference trajectory for the controlled output variable 
$\bm{y}(t)$ is denoted as $\bm{r}(t)$. $\mathbf{Q_s}$ and 
$\mathbf{Q_t}$ are weight matrices of appropriate dimension, weighting the 
contribution of the \textit{stage} tracking cost and the \textit{terminal} 
tracking cost, respectively. In this notation, 
$\| \bm{a} \|^2_\mathbf{A} := \bm{a}^\top \mathbf{A} \bm{a}$ denotes the
squared norm of a vector $\bm{a}$ weighted by the matrix $\mathbf{A}$.

\section{Solution strategy}
\label{sec:solution_strategy}
As mentioned in the Introduction, Problem~\eqref{eq:sw-opt} can in principle be viewed through the lens of \textit{mixed-integer dynamic optimization}. If each forcing period is divided into $n_g$ fine-grid subintervals, the formulation introduces binary variables $z_{i,k,j}\in\{0,1\}$ and the total number of binary decision variables scales with $n_u n_T n_g$. The PWM ON-OFF structure can be enforced by monotonicity constraints, e.g., $z_{i,k,j}\geq z_{i,k,j+1}$, but finer grids are needed for accurate switching-time approximation, significantly increasing the size of the resulting optimization problem.

Here, we propose an RL-based solution approach. The key idea is \textit{not} to optimize the piecewise binary inputs over a fine time grid explicitly, but rather to work with duty cycles as continuous proxy variables that encode the underlying binary inputs while preserving the ON-to-OFF structure within each forcing period. This avoids fine-grid binary decision variables and yields an optimization problem compatible with continuous \textit{action} spaces.

In $\mathcal{T}_k$, the duty cycle of input $i$ (cf. Eq.~\eqref{eq:binary_input}), 
$D_{i,k} \in [0,1]$, is defined as 
the time average of the binary input:
\begin{equation}
    \begin{aligned}
    D_{i,k} 
    &= \frac{1}{T} \int_{kT}^{(k+1)T} u_{i,k}(t)\,dt 
     = \frac{1}{T} \int_{kT}^{\tau_{i,k}} 1\,dt
      = \frac{\tau_{i,k}-kT}{T}, \\
    &\forall k \in \{0, 1, \dots, n_T-1\}, \,\forall i \in \{1, \dots, n_u\}.
    \label{eq:duty}
    \end{aligned}
\end{equation}

As a consequence, $\tau_{i,k}$ is uniquely determined by $D_{i,k}$ via:
\begin{equation}
    \tau_{i,k}(D_{i,k}) = \bigl(k + D_{i,k}\bigr)T.
\end{equation}

Let $\bm{D}_k \in [0,1]^{n_u}$ collect the duty cycles of all inputs in forcing period $k$, 
$\bm{\tau}_k \in [kT,(k+1)T]^{n_u}$ the corresponding switching times, $\bm{u}_k(t) := [u_{1,k}(t),\dots,u_{n_u,k}(t)]^\top \in \{0,1\}^{n_u}$ the time-varying binary input vector over $\mathcal{T}_k$, and $\bm{I}_k(t) := [I_{1,k}(t),\dots,I_{n_u,k}(t)]^\top \in \mathbb{R}^{n_u}$ the corresponding physical light intensity. We treat $\bm{D}_k$ as the decision variable over $\mathcal{T}_k$ since $\bm{D}_k \mapsto \bm{\tau}_k(\bm{D}_k) 
\mapsto \bm{u}_k(t) \mapsto \bm{I}_k(t)$, which ultimately drives the system dynamics.

Furthermore, we describe the state transition (cf. Eq.~\eqref{eq:dynamics}) as a Markov decision process:
\begin{equation}
    \bm{x}_{k+1} \sim \mathrm{p}\bigl(\cdot \mid \bm{x}_k, \bm{D}_k\bigr),
    \, k \in \{0, 1, \dots, n_T-1\},
\end{equation}
where $\mathrm{p}$ denotes a probability density function.

Let $\pi(\bm{D}_k \mid \bm{x}_k, \bm{\theta})$ be the control policy, parametrized by $\bm{\theta} \in \mathbb{R}^{n_\theta}$. We transform the optimization problem \eqref{eq:sw-opt} to:
\begin{equation} 
    \max_{\bm{\theta}} \; \mathbb{E}_{\bm{\tau}\sim\mathrm{p}(\bm{\tau}\mid\bm{\theta})} \left[ J(\bm{\tau}) \right], 
    \label{eq:rl_problem} 
\end{equation} 
where $\bm{\tau}$ is a trajectory of states, inputs, rewards, and transitions:
\begin{equation} 
    \bm{\tau} = \{ (\bm{x}_k, \bm{D}_k, r_{k+1}, \bm{x}_{k+1}) \}_{k=0}^{n_T-1},
    \label{eq:joint_tau}
\end{equation}
and the reward at time $k$ is defined as:
\begin{subequations}
    \begin{align}
        r_{k} &:= -\| \bm{y}_k - \bm{r}_{k} \|^2_{\mathbf{Q_s^*}}, 
        \quad k \in \{1, \dots, n_T-1\}, \\
        r_{n_T} &:= -\| \bm{y}_{n_T} - \bm{r}_{n_T} \|^2_{\mathbf{Q_t^*}}.
    \end{align}
\end{subequations}
This mirrors Eq.~\eqref{eq:cost}: intermediate negative tracking errors (stage rewards) are weighted by $\mathbf{Q_s^*}$, while the negative final-time tracking error (terminal reward) is weighted by $\mathbf{Q_t^*}$. Consistent with Eq.~\eqref{eq:joint_tau}, we collect the first reward after the first duty-cycle action has been applied.

The objective is then redefined in the RL formulation as the return:
\begin{equation}
    J := \sum_{k=1}^{n_T} r_{k},
\end{equation}
which corresponds to an \textit{undiscounted episodic return}. It is therefore the discrete-time counterpart of the objective in Eq.~\eqref{eq:cost}, using negative costs so that maximizing $J$ in \eqref{eq:rl_problem} corresponds to minimizing Eq.~\eqref{eq:cost}.

To solve \eqref{eq:rl_problem}, we apply policy gradients in a gradient-ascent fashion over $n_m$ epochs, with learning rate $\alpha$:
\begin{equation}
    \bm{\theta}_{m+1} = \bm{\theta}_m + \alpha \nabla_{\bm{\theta}} \, \mathbb{E}_{\bm{\tau}} \left[ J(\bm{\tau}) \right],
    \quad m = 0, \dots, n_m-2.
\label{eq:update_rule_general}
\end{equation}

The joint probability density of the trajectory $\bm{\tau}$ follows:
\begin{equation}
\mathrm{p}(\bm{\tau} \mid \bm{\theta}) 
= \mathrm{p}(\bm{x}_0) 
\,
\prod_{k=0}^{n_T-1} 
\left[ 
    \pi(\bm{D}_k \mid \bm{x}_k, \bm{\theta}) 
    \, 
    \mathrm{p}(\bm{x}_{k+1} \mid \bm{x}_k, \bm{D}_k) 
\right].
\label{eq:prob_dis_tau}
\end{equation}

Applying the Policy Gradient Theorem \citep{NIPS1999_464d828b} and approximating the expectation via $n_\mathrm{MC}$ Monte Carlo simulations leads to:
\begin{equation} 
\begin{aligned} \nabla_{\bm{\theta}} \mathbb{E}_{\bm{\tau}} \left[ J(\bm{\tau}) \right] &= \mathbb{E}_{\bm{\tau}} \left[ J(\bm{\tau}) \nabla_{\bm{\theta}} \left[\sum_{k=0}^{n_T-1} \log \pi(\bm{D}_k \mid \bm{x}_k, \bm{\theta}) \right] \right], \\ &\approx \frac{1}{n_\mathrm{MC}} \sum_{j=1}^{n_\mathrm{MC}} \left[ \left( \frac{J\left( \bm{\tau}^{(j)} \right) - \bar{J}_{m}}{\sigma_{J_{m}} + \epsilon} \right) \right. \\ &\quad \left. \cdot \nabla_{\bm{\theta}} \left[ \sum_{k=0}^{n_T-1} \log\left( \pi(\bm{D}_k^{(j)} \mid \bm{x}_k^{(j)}, \bm{\theta}) \right) \right] \right],
\end{aligned} \label{eq:grad_exp_2} 
\end{equation}
with a baseline normalized by the mean $\bar{J}_{m}$ and the standard deviation $\sigma_{J_{m}}$ of the return in epoch $m$; $\epsilon$ is a small positive scalar (\textit{machine epsilon}).

\section{PWM-driven control of optogenetic cell growth}
\label{sec:case_study}
We consider an \textit{E. coli} strain with an engineered lysine auxotrophy obtained by deletion of \textit{lysA} (encoding diaminopimelate decarboxylase), which is essential for lysine biosynthesis. 

\subsection{Mathematical model}
\label{subsec:model}
The corresponding macroscopic bioprocess dynamics for optogenetic growth in a well-mixed chemostat with dilution rate $d_l$ follow \citep{espinel-rios_reinforcement_2025}:
\begin{subequations}
\begin{align}
    \odv{b}{t} &= \bigl(\mu(g,p) - d_l\bigr) \, b, \, b(0)=b_0,\\
    \odv{g}{t} &= - q_{g}(g,p)\, b + (g_{\text{in}} - g) \, d_l, \, g(0)=g_0, \label{eq:s_ode}\\
    \odv{p}{t} &= q_{p}(I) - \bigl(d_{p} + \mu(g,p)\bigr) \, p, \, p(0)=p_0,  \label{eq:e_i_ode}
\end{align}
\end{subequations}
with kinetic rates:
\begin{subequations}
\begin{align}
    \mu(g,p) &= \mu_{\text{max}} 
    \left( \frac{g}{g + k_{g}} \right) 
    \left( \frac{ f_{c} p}{ f_{c} p + k_{p}} \right),\label{eq:mu_i}\\
    q_{g}(g,p) &= Y_{g/b} \, \mu(g,p),  \label{eq:q_s_i}\\
    q_{p}(I) &= q_{p,\text{max}} \left( \frac{I^{n}}{I^{n} + k_{I}^{n}} \right),\label{eq:q_e_i}
\end{align}
\end{subequations}
where $b \in \mathbb{R}$ is the biomass concentration, $g \in \mathbb{R}$ the extracellular glucose concentration, and $p \in \mathbb{R}$ the intracellular lysine concentration. $I \in \mathbb{R}$ is the light intensity. The intensity-dependent lysine synthesis rate $q_p(I) : \mathbb{R} \mapsto \mathbb{R}$ is a Hill function that lumps the light-controlled expression of \textit{lysA} together with the resulting lysine production. The parameter $d_p$ accounts for lysine conversion and usage to support growth and viability.

We assume a ccaS/ccaR two-component optogenetic system \citep{davidson_programming_2013}, which is inducible by green light. We assume the following parameters: $n = 0.2191$ and $k_I = 5.5086\times 10^{-7} \, \mathrm{W/m^2}$, adapted and fitted (without leakage) from the dose-response data in \cite{davidson_programming_2013}, and $q_{p,\text{max}} = 0.3366 \, \mathrm{mmol/(g \cdot h)}$ (from \cite{espinel-rios_reinforcement_2025}). Light intensity is measured in $\mathrm{W/m^2}$. The remaining parameters are \citep{espinel-rios_reinforcement_2025}: 
$\mu_{\text{max}} = 0.982 \, \mathrm{h^{-1}}$,
$f_{c}=1100 \, \mathrm{g/L}$, 
$Y_{g/b}=10.18 \, \mathrm{mmol/g}$, 
$k_{g}=2.964 \times 10^{-4} \, \mathrm{mmol/L}$, 
$k_{p} = 1.7 \, \mathrm{mmol/L}$, 
$d_l=0.15 \, \mathrm{h^{-1}}$, 
$g_{\text{in}}=200 \, \mathrm{mmol/L}$, 
$d_p = 20.8 \, \mathrm{h^{-1}}$.

\subsection{Smoothing dose-response curves via duty cycles}
\label{sec:intensity_vs_duty_cycle}
Before diving into the implementation of the RL strategy, we illustrate the smoothing effect of PWM on steep dose-response curves via duty-cycle averaging. Let us consider a binary ON-OFF input sequence over the forcing period $\mathcal{T}_k$ and one light-intensity channel (green light) ($n_u = 1$), with $I_{1,\max} = 30\,\mathrm{W/m^2}$ and duty cycle $D_{1,k} \in [0,1]$. Hereafter, we omit the subscript indicating input channel 1 for simplicity, as there is only one input channel in the process ($I_{1,\max}:=I_{\max}$, $D_{1,k}:=D_{k}$). During the ON subinterval, the intensity is $I(t) = I_{\max}$, and during the OFF subinterval it is $I(t) = 0$ (cf. Eq. \eqref{eq:physical_i_I}).

\begin{figure}[htb!]
\begin{center}
\includegraphics[scale=0.5]{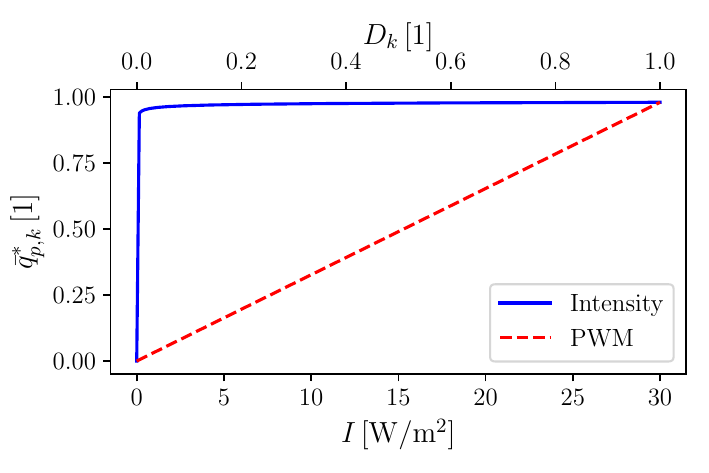}
\vspace{-0.4cm}
\caption{Comparison of the \textit{normalized} average Hill activation function $\bar{q}_{p,k}^*$ over the forcing period $\mathcal{T}_k$ under intensity-driven and PWM-driven actuation. The normalized average activation is defined as $\bar{q}_{p,k}^* := \bar{q}_{p,k}/q_{p,\text{max}}$, yielding a range $[0,1]$.}
\label{fig:intensity_vs_duty_cycle}
\end{center}
\end{figure}

The period-averaged value of the Hill function, denoted as $\bar{q}_p$, over the interval $[kT, (k+1)T]$ is:
\begin{equation}
\begin{aligned}
\bar{q}_{p,k} 
&= \frac{1}{T}
\int_{kT}^{(k+1)T} q_{p}(I(t)) \, \mathrm{d}t \\
&= \frac{1}{T}\left[
\int_{kT}^{\tau_{k}=(k + D_k)T} q_{p}(I_{\max}) \, \mathrm{d}t
+
\int_{\tau_{k}=(k + D_k)T}^{(k+1)T} q_{p}(0) \, \mathrm{d}t
\right] \\
&= D_k\, q_{p}(I_{\max}).
\label{eq:duty_avg}  
\end{aligned}
\end{equation}
Thus, under the no-leakage assumption in Eq.~\eqref{eq:q_e_i}, the average Hill activation over the forcing period depends linearly on the duty cycle $D_k$.

The parameters of the Hill function introduced in the previous section yield a very steep intensity-response curve. Fig.~\ref{fig:intensity_vs_duty_cycle} therefore illustrates how duty-cycle modulation provides a smoother \textit{period-averaged} input-output map than direct intensity modulation, thereby facilitating practical tunability and reinforcing the advantages of PWM control.

\subsection{RL-derived control policies}
To evaluate the proposed RL-based solution to the switching-time optimal control problem, we consider a three-setpoint biomass reference trajectory involving transitions from $3\,\mathrm{g/L}$ to $5\,\mathrm{g/L}$ and finally to $7\,\mathrm{g/L}$ over $24\,\mathrm{h}$, with 24 forcing periods of $1\,\mathrm{h}$ each. Such a trajectory may arise in synthetic microbial consortia where biomass concentration is linked to the productivity of a specific metabolic submodule carried by a specific strain. The corresponding control problem is to find the 24 duty cycles (encoding the ON-to-OFF switching times) for the 24 forcing periods that best track this reference trajectory.

We now introduce our control scenarios:
\begin{itemize}
    \item \textbf{Ideal system}: we assume no uncertainty in the process dynamics. 
    \item \textbf{Uncertain system}: we assume different uncertainty levels in the process dynamics. Specifically, we assume that the initial conditions of the plant and the maximum gene expression rate $q_{p,\text{max}}$ follow normal distributions with standard deviations of $2.5, 5,$ and $7.5\,\%$ relative to their mean (nominal) values.
\end{itemize}

We use the model in Section \ref{subsec:model} as a \textit{digital twin} for our optogenetic bioprocess, and thus as the virtual environment to train our RL policies. The nominal initial conditions are $b_0 = 3 \,\mathrm{g/L}$, $g_0 = 50 \,\mathrm{mmol/L}$, and $p_0 = 1.0752\times 10^{-4} \,\mathrm{mmol/g}$.

\begin{figure}[h!]
	\begin{center}
        \subfigure[\textbf{Intensity; 0 \% unc.}]{\includegraphics[scale=0.36, trim={0 0 0 0}, clip]{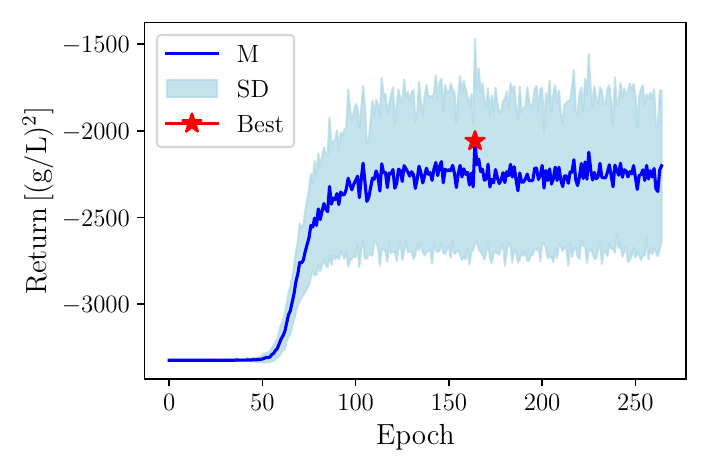}}
        \subfigure[\textbf{PWM; 0 \% unc.}]{\includegraphics[scale=0.36, trim={0 0 0 0}, clip]{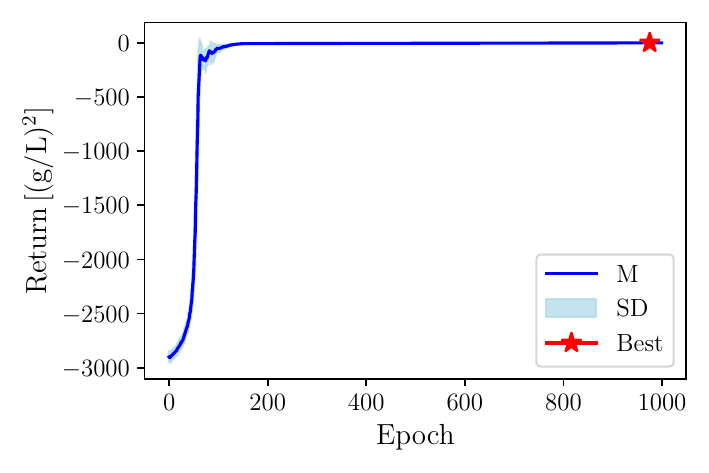}}
        \subfigure[\textbf{PWM; 2.5 \% unc.}]{\includegraphics[scale=0.36, trim={0 0 0 0}, clip]{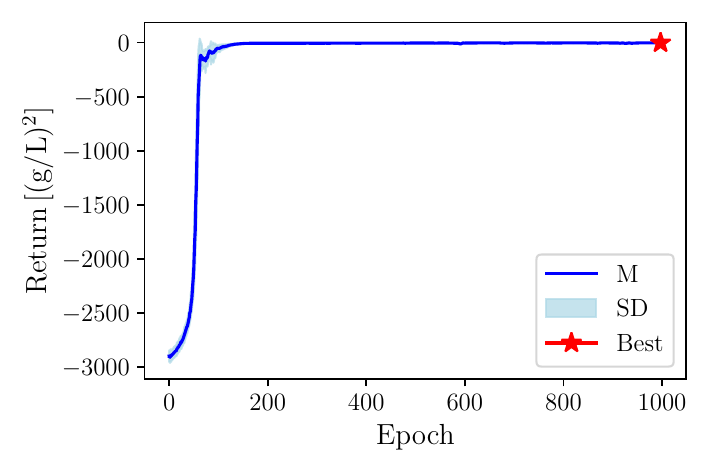}}
        \subfigure[\textbf{PWM; 5 \% unc.}]{\includegraphics[scale=0.36, trim={0 0 0 0}, clip]{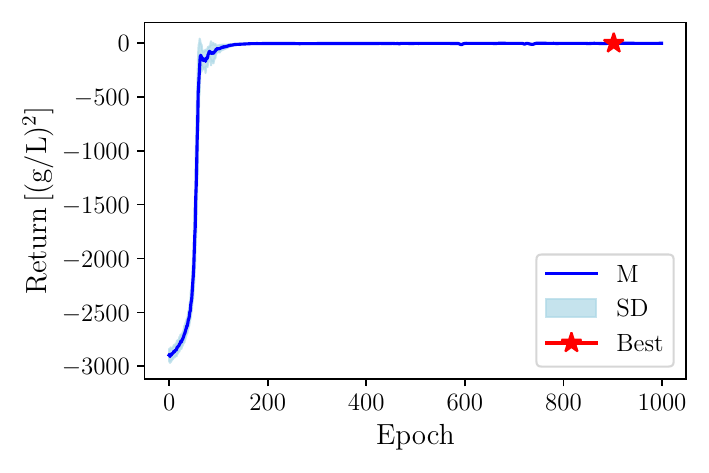}}
        \subfigure[\textbf{PWM; 7.5 \% unc.}]{\includegraphics[scale=0.36, trim={0 0 0 0}, clip]{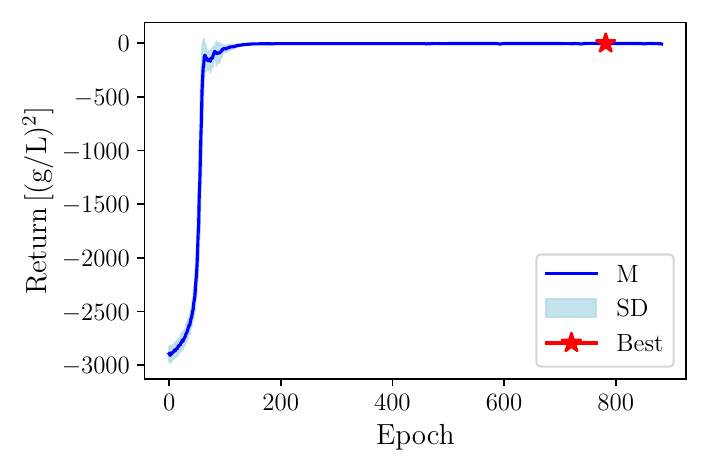}}
        \vspace{-0.3cm}
		\caption{Return (negative tracking error) over training epochs for the RL policies under intensity-driven and PWM-driven optogenetic control for selected uncertainty (unc.) levels. M: mean; SD: standard deviation; Best: selected policy.}
		\label{fig:rl_return_policies}
	\end{center}
\end{figure}

\begin{figure}[h!]
	\begin{center}
        \subfigure[\textbf{Intensity; 0 \% uncertainty}]{\includegraphics[scale=0.6, trim={0 0 0 0}, clip]{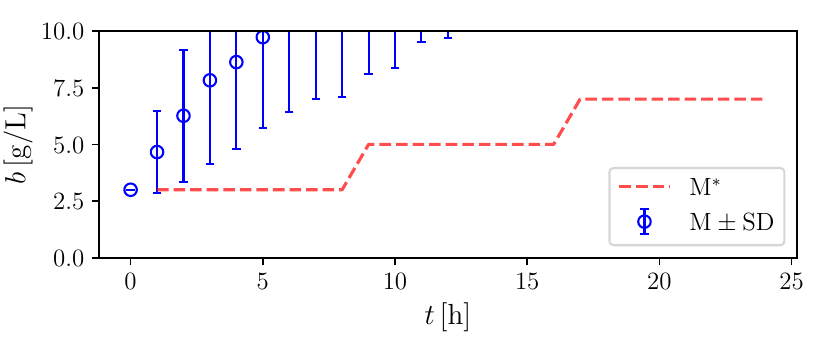}}\\[-0.1cm]
        \subfigure[\textbf{PWM; 0 \% uncertainty}]{\includegraphics[scale=0.6, trim={0 0 0 0}, clip]{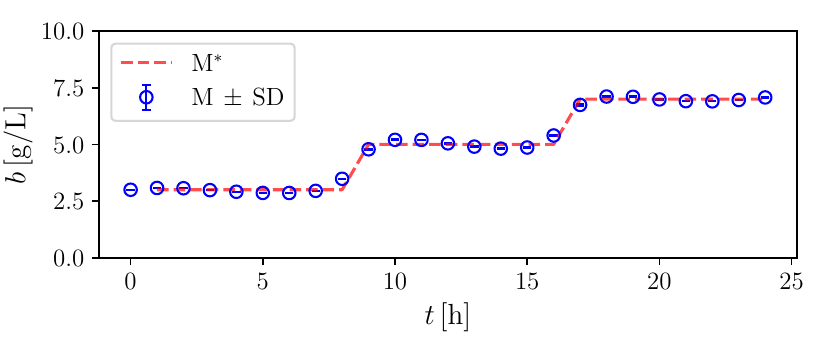}}\\[-0.1cm]
        \subfigure[\textbf{PWM; 2.5 \% uncertainty}]{\includegraphics[scale=0.6, trim={0 0 0 0}, clip]{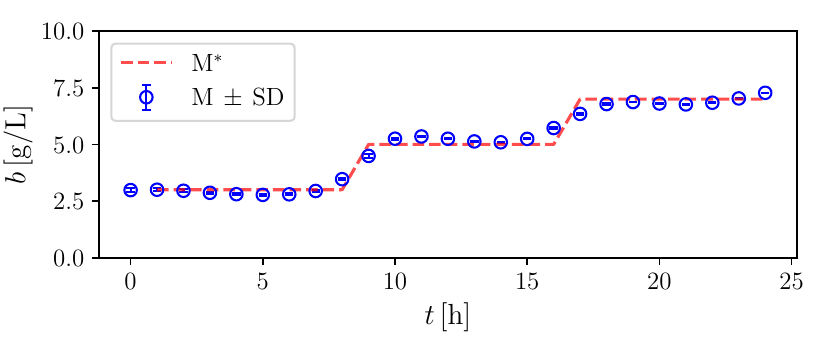}}\\[-0.1cm]
        \subfigure[\textbf{PWM; 5 \% uncertainty}]{\includegraphics[scale=0.6, trim={0 0 0 0}, clip]{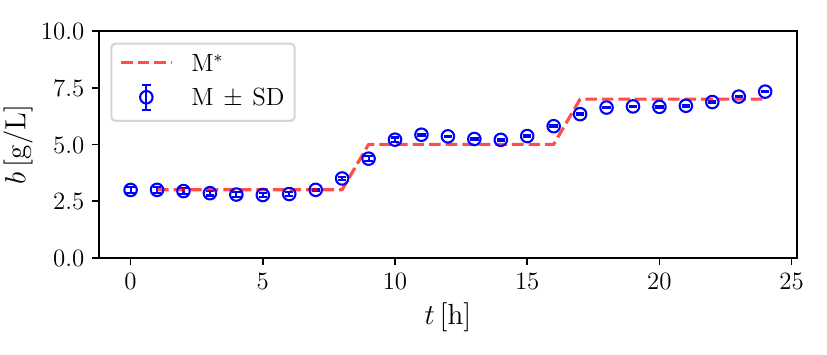}}\\[-0.1cm]
        \subfigure[\textbf{PWM; 7.5 \% uncertainty}]{\includegraphics[scale=0.6, trim={0 0 0 0}, clip]{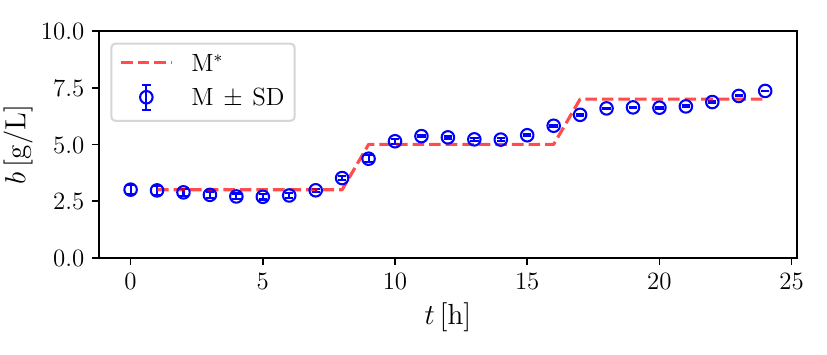}}\\[-0.1cm]
        \vspace{-0.3cm}
		\caption{Optimized biomass trajectories versus the three-setpoint reference trajectory obtained with RL policies under intensity-driven and PWM-driven optogenetic control for selected uncertainty levels. M: mean; SD: standard deviation; M$^*$: reference.}
		\label{fig:setpoint_tracking}
	\end{center}
\end{figure}

\begin{figure}[h!]
	\begin{center}
        \subfigure[\textbf{Intensity; 0 \% uncertainty}]{\includegraphics[scale=0.59, trim={29 0 0 0}, clip]{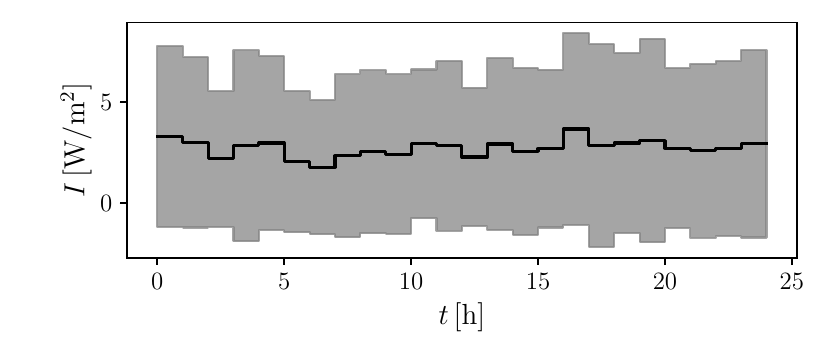}}\\[-0.1cm]
        \subfigure[\textbf{PWM; 0 \% uncertainty}]{\includegraphics[scale=0.59, trim={0 0 0 0}, clip]{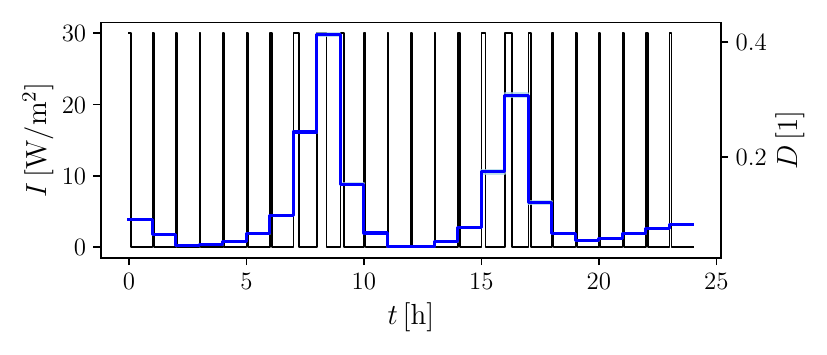}}\\[-0.1cm]
        \subfigure[\textbf{PWM; 2.5 \% uncertainty}]{\includegraphics[scale=0.59, trim={0 0 0 0}, clip]{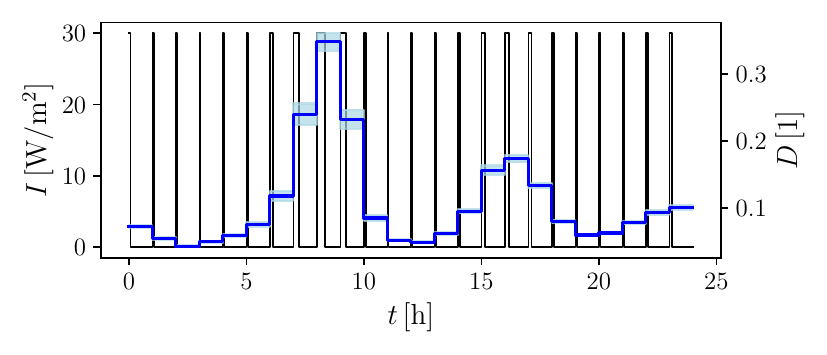}}\\[-0.1cm]
        \subfigure[\textbf{PWM; 5 \% uncertainty}]{\includegraphics[scale=0.59, trim={0 0 0 0}, clip]{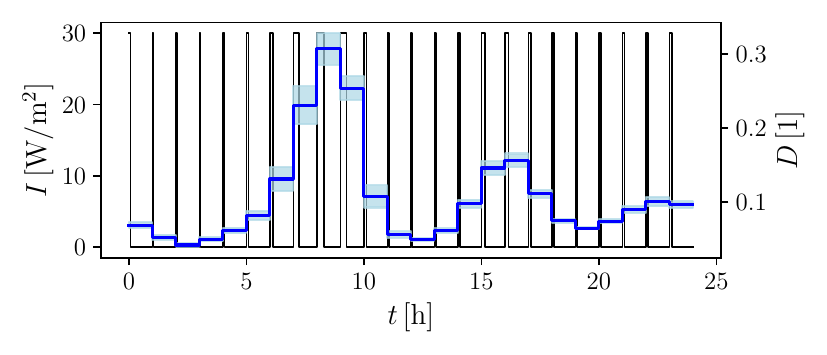}}\\[-0.1cm]
        \subfigure[\textbf{PWM; 7.5 \% uncertainty}]{\includegraphics[scale=0.59, trim={0 0 0 0}, clip]{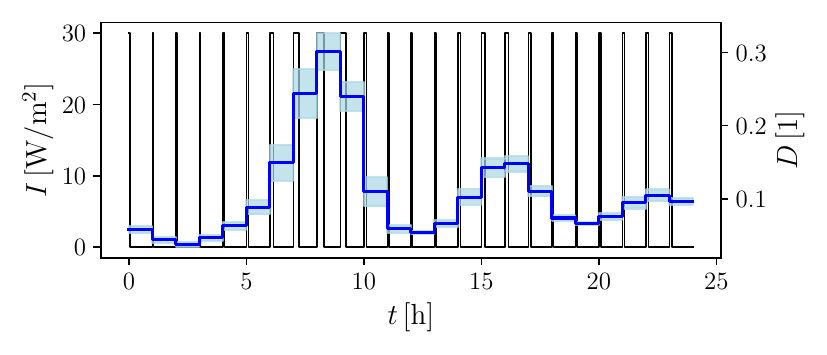}}\\[-0.1cm]
    
    \vspace{-0.4cm}
    \caption{Optimized light input trajectories obtained with RL policies under intensity-driven and PWM-driven optogenetic control for selected uncertainty levels. The duty cycle mean is denoted by a blue line; the physical light intensity mean is denoted by a black line. In (a), the standard deviation of the intensity is represented as a shaded gray area. In (b)-(e), the standard deviation of the duty cycle is denoted as a shaded blue area.}
		\label{fig:rl_input_duty_intensity}
	\end{center}
\end{figure}

We train RL policies as outlined in Section \ref{sec:solution_strategy} using PyTorch \citep{pytorch}. We use a fully connected stochastic policy network (4 hidden layers, 20 neurons each, Leaky ReLU). An output linear layer returns the mean and standard deviation of a normal distribution over the duty cycle. We use $n_m = 1000$, $n_\mathrm{MC} = 100$, and $\alpha = 0.001$. Early stopping is applied if no improvement in the return is observed for 100 consecutive epochs. To facilitate convergence, we augment the policy input with the current and previous states and duty cycles, as well as a linear process-time embedding $e_k \in [-1, 1]$. Accordingly, the policy is reformulated as $\pi\bigl(D_k \mid \bm{x}_k, \bm{\theta}\bigr) 
    := \pi\bigl(D_k \mid \bm{z}_k, \bm{\theta}\bigr)$, where $\bm{z}_k 
    := \bigl[ \bm{x}_k^\top,\; D_{k-1},\; \bm{x}_{k-1}^\top,\; D_{k-2},\; e_k \bigr]^\top$. Note that this procedure corresponds to the \textit{in-silico} training setup. Once trained, deployment of the policy only requires a forward evaluation of the neural network to compute the duty cycle, which is computationally inexpensive and compatible with real-time bioprocess operation.

Fig.~\ref{fig:rl_return_policies} shows the evolution of the return over training epochs, together with the selected (best) policy. As a benchmark, we also trained a policy based directly on light intensity: here, the intensity was allowed to vary in the range $[0, I_{\max}]$ as a piecewise-constant input with the same interval length as the forcing periods in the PWM cases. Even in the absence of system uncertainty, the intensity-driven control scenario resulted in very low returns with large variability, as reflected by the large standard deviation of the return (cf. Fig.~\ref{fig:rl_return_policies}a). This can be explained by the extremely steep Hill-type activation $q_p(I)$ (cf. Fig.~\ref{fig:intensity_vs_duty_cycle}), which induces almost binary ON-OFF gene-expression transitions and thus yields weakly informative gradients for learning. This effectively stalls the learning process, even though, \textit{theoretically} speaking, intensity actuation does cover the range $[0, q_{p,\text{max}}]$, despite the steepness. Note that we do not plot the intensity-driven benchmark for the uncertainty scenarios, as it failed to converge even in the most optimistic case without uncertainty.

In contrast, the duty-cycle control, both without and with system uncertainty, led in all cases to high return values and clear convergence of the learning process (cf. Fig.~\ref{fig:rl_return_policies} (b)-(e)), demonstrating the effectiveness and robustness of the proposed PWM-driven RL approach. The smoother tunability provided by PWM allows the policy to experience more stable and informative gradients during learning. The converged PWM-driven policies also exhibited low variability, which can be explained by the fact that, in essence, the duty cycle operates between only two modes (light ON/OFF within each forcing period). This makes the process less sensitive to uncertainties in intermediate gene activation levels.

The poor convergence of the intensity-driven optogenetic bioprocess is clearly reflected in Fig. \ref{fig:setpoint_tracking}, which compares the optimized biomass trajectories against the three-setpoint reference trajectory. In contrast, the PWM-driven bioprocess achieves successful tracking in all cases (with and without uncertainty), closely matching the reference. As expected, the PWM-driven process with no uncertainty yields practically perfect reference tracking, consistent with the return convergence behavior discussed above. Moreover, the low variability in the converged returns translates into very low variability in the biomass dynamics, even at the largest uncertainty level. As already mentioned, this is related to the fact that the process effectively operates between two robust modes (light ON/OFF within each forcing period), which are expected to be less sensitive to system uncertainty than intermediate gene-expression levels (as opposed to intensity-driven control). That said, as uncertainty increases, PWM-based control begins to struggle slightly at the setpoint transitions, yet the overall tracking performance remains within acceptable margins.

Finally, taking a closer look at the optimized input trajectories (Fig. \ref{fig:rl_input_duty_intensity}), we see that the intensity-driven control remains practically constant in terms of its mean intensity, but with a very large standard deviation, which is consistent with the poor learning convergence discussed above. In contrast, the PWM-driven control scenarios effectively modulate the duty cycles to match the reference trajectory. The duty cycles follow a coherent pattern across forcing periods, oscillating from lower to higher values to both maintain setpoints and enable setpoint transitions when appropriate. There is some increased standard deviation in the duty-cycle trajectories with rising uncertainty, yet it does not grow significantly. Overall, this demonstrates that our RL approach can robustly learn continuous duty cycles that encode binary ON-OFF intensity patterns within forcing periods, resulting in successful PWM-driven control policies in the context of optogenetic bioprocesses.

\section{Conclusion}
\label{sec:conclusion}
\sloppy
In this study, we introduced an RL strategy to solve switching-time optimal control problems tailored to PWM optogenetics. The approach leverages duty cycles as continuous decision variables, which are decoded into binary ON-OFF light-intensity profiles over forcing periods. On this basis, we use policy gradients to optimize the control policy. A case study involving optogenetic growth control in the presence of a steep light-gene-expression dose-response demonstrated the strong performance of the RL-derived PWM control policies, both in terms of tunability of the controlled variable and reference tracking under different levels of parametric uncertainty. The conducted robustness analysis was restricted to uncertainty distributions incorporated during policy training, with the aim of demonstrating that effective policies could still be learned in uncertain environments. This setting is also compatible with domain-randomization strategies, provided that plausible uncertainty ranges can be specified. Future work should assess robustness to out-of-distribution scenarios, such as process disturbances and unmodeled dynamics not considered during training, as well as policy adaptability when transferred to experimental bioprocesses.

Beyond optogenetics, the proposed control approach can be generalized to other bioprocesses involving duty-cycle-type inputs, such as microalgal photocycles or pulsed/intermittent feeding. Ongoing work focuses on extending the methodology to more complex scenarios, including multiple-input systems and simultaneous optimization of forcing periods and duty cycles.

\bibliography{ifacconf}

\end{document}